\documentclass[12pt]{article}
\usepackage{a4,amsthm,amsfonts,
}
\swapnumbers 
\theoremstyle{plain}
\newtheorem{thm}{THEOREM}[section]
\newtheorem{theorem}{THEOREM}[section]

\newtheorem{lm}[thm]{LEMMA}

\newtheorem{proposition}[thm]{PROPOSITION}
\newcommand{\RR}{\mathord{\mathbb R}}
\newcommand{\CC}{\mathord{\mathbb C}}
\newcommand{\NN}{\mathord{\mathbb N}}

\newcommand{\HH}{{\cal H}}
\newcommand{\MM}{{\cal M}}
\newcommand{\OO}{{\cal O}}
\def\mfr#1/#2{\hbox{${{#1} \over {#2}}$}}
\def\const.{{\rm const.}}
\newcommand{\beq}{\begin{equation}}
\newcommand{\eeq}{\end{equation}}
\def\beqa{\begin{eqnarray}}
\def\eeqa{\end{eqnarray}}

\newcommand{\text}{\hbox}

\def\mfr#1/#2{\hbox{${{#1} \over {#2}}$}}

\begin{document}
\title{\bf{
Geometric Modular Action,\\
Wedge Duality and Lorentz Covariance\\are Equivalent\\
for Generalized Free Fields}}
\author{\vspace{5pt}Johanna Gaier$^{1}$ and 
        Jakob Yngvason$^{2}$\\
      \vspace{-4pt}\small{$1.$ Abteilung f\"ur Finanz- und 
      Versicherungsmathematik, TU 
Wien,}\\
\small{Wiedener Hauptstra\ss e 8--10, A-1040 Vienna, Austria}\\  
        \vspace{-4pt}\small{$2.$ Institut f\"ur Theoretische Physik, 
Universit\"at 
Wien}\\
\small{Boltzmanngasse 5, A-1090 Vienna, Austria}\\
        }

\maketitle

\newcommand{\I}{1 \hspace{-1.1mm} {\rm I}}
\newcommand{\AO}{${\cal A}$(${\cal O}$)\ }
\newcommand{\AG}{${\cal A}$(${\cal G}$)\ }

\begin{abstract}

The Tomita-Takesaki modular groups and conjugations for the observable
algebras of space-like wedges and the vacuum state are computed for
translationally covariant, but possibly not Lorentz covariant,
generalized free quantum fields in arbitrary space-time dimension $d$.
It is shown that for $d\geq 4$ the condition of
geometric modular action (CGMA) of Buchholz, Dreyer, Florig and
Summers \cite{BDFS}, Lorentz covariance and wedge duality are all
equivalent in these models.  The same holds for $d= 3$ if there is
a mass gap.  For massless fields in $d=3$, and for $d=2$ and arbitrary
mass, CGMA does not imply Lorentz covariance of the field itself, but
only of the maximal local net generated by the field.

\end{abstract}

\section{Introduction}

The importance of Tomita-Takesaki modular theory for both structural
analysis and constructive aspects of quantum field theory has been amply
manifested by important publications in recent years.  We refer to
\cite{B1},\cite{Sch} and \cite{BFS} for extensive lists of references on
this subject.  The Bisognano-Wichmann theorem (\cite{BW1}, \cite{BW2}),
proved already in 1975, is the basic insight on which these developments
are founded.  It provides a geometrical interpretation of the modular objects
associated with algebras generated by Poincar\'e covariant Wightman field
operators localized in space-like wedges.

In 1992 Borchers \cite{B2} discovered an important partial converse to this
theorem. He showed that in two space-time dimensions the modular objects
associated with a translationally covariant local net of von Neumann
algebras and a vacuum state lead to a representation of the Poincar\'e
group, even if no Lorentz covariance of the net is required at the outset.
(See also \cite{F} for a simplified proof of Borchers' theorem.)  Such a
geometrical interpretation of the modular objects is not always possible in
higher dimensions, however, as can be seen from examples given in \cite{Y}.

By {\it postulating} a certain form of geometric action of the modular
conjugations associated with space-like wedges and a given state (\lq\lq
Condition of geometric modular action") Buchholz et al.  \cite{BDFS} were able
to construct a representation of the Poincar\'e group in space-time dimension 4
without even assuming translational invariance.  The essence of the CGMA is the 
requirement that the modular conjugation of every wedge leaves the family of all 
wedge algebras invariant. As shown in \cite{BFS} the
spectrum condition for the translations follows from the additional requirement
that the group generated by the conjugations contains the modular groups of the
wedge algebras (\lq\lq Modular stability condition").  Such a purely algebraic
characterization of vacuum states has the potential for generalizations to a
stability condition for quantum fields on curved space-times.  Other important 
results relying on geometric actions of modular groups have been obtained, 
e.g., 
in \cite{BGL}, \cite{G}, \cite{GL}.

As a contribution to the understanding of the possible modular actions in 
quantum
field theory when the Bisognano-Wichmann theorem does not apply we compute in
this note the modular groups and the modular conjugations associated with the
wedge algebras generated by translationally covariant generalized free quantum
fields in arbitrary space-time dimension $d$.  Such a computation was carried 
out
in \cite{Y} for two dimensional fields depending only on one light
coordinate, and certain special cases in higher dimensions.  Here we treat the
general case (for single component, hermitian fields).

We investigate the geometrical
significance of the modular objects, and in particular we answer the
question when the adjoint action of the modular conjugation associated with
a wedge algebra leaves the set of all wedge algebras invariant.  We show that 
in $d\geq 4$
this is the case if and only if the two point function defining the field
is Lorentz invariant.  In fact, Lorentz invariance follows already from wedge 
duality for the field, i.e., if the algebra of a wedge is the commutant of the 
algebra of the opposite wedge, which is a consequence of  CGMA cf.\ Prop.\ 
4.3.1 
in \cite{BDFS}. Besides the explicit formulas for the modular
objects this result is based on a lemma concerning the zeros of polynomials
restricted to a mass shell (Lemma \ref{basic}). The same conclusion can be 
drawn 
from the requirement that the modular groups act locally, i.e., transform 
observables localized in a bounded region into observables localized in another 
bounded region.

In view of the general result of \cite{BDFS} the Lorentz covariance of fields
satisfying CGMA is not a surprise, but it is important to note that this is not 
a
consequence of \cite{BDFS} alone.  The point is that the same wedge algebras
could {\it a priori} be generated by different fields and not all of them might
be Lorentz covariant.  In fact, the wedge algebras for a massless
free field in $d=3$ can be generated by certain derivatives of the field that 
do
break Lorentz invariance.  In $d=3$, however, this massless case is the only
exception:  if there is a mass gap, then CGMA implies Lorentz covariance of the
field.  In $d=2$ also massive fields without Lorentz covariance can fulfill 
CGMA.
In the cases where Lorentz covariance of the field is broken but CGMA holds the
minimal local net generated by the field operators is not Lorentz covariant, in
contrast to the maximal net defined by intersections of wedge algebras which is
strictly larger in these cases. 

The bottom line is that in $d\geq 4$  the following conditions are all 
equivalent 
for the  models considered:  a) CGMA, b) 
wedge duality, c) local action of the modular groups d) Lorentz covariance of 
the field. This equivalence holds also for $d=3$, provided there is a mass gap.

\section{Definition of the Models}

We consider a Hermitian Wightman field $\Phi$ that transforms 
covariantly under space-time 
translations, but not necessarily under Lorentz transformations.
The general structure of the 2-point-function ${\cal W}_2(x-y)=\langle\Omega,
\Phi(x)\Phi(y)\Omega\rangle$, where $\Omega$ denotes the vacuum state, follows 
from the
Jost-Lehmann-Dyson re\-pre\-sen\-ta\-tion 
 (cf.\ e.g. \cite{B3}); its Fourier transform can be written
\beq
\widetilde{\cal W}_{2}(p) = \int_{0}^{\infty} M(p, m^{2}) 
\Theta(p_{0})\delta(p^{2} - m^{2}) 
\,{\rm d}\rho(m^{2})\label{c}
\eeq
where the Lehmann weight ${\rm d}\rho$ is a positive, tempered measure on 
$\RR_+$ 
and  $M(p, m^{2})$ is for fixed $m$ an even polynomial in 
$p=(p_0,\dots,p_{d-1})\in\RR^d$, i.e., 
\beq
M(p, m^{2})=M(-p, m^{2}),
\eeq
 with
\beq
M(p, m^{2})\geq 0\quad{\rm for}\quad p\in H_{m}^+ := \{ p\in \RR^d: 
p^{2} - m^{2} = 0, p_0\geq 0 \}
\eeq
and ${\rm d}\rho$-almost all $m^2$.

The Hilbert space of the field is the symmetric Fock space over the \lq\lq 
one-field-space\rq\rq\  ${\cal H}^{(1)}$, which is the $L^2$ space 
corresponding 
to the positive measure 
$\widetilde{\cal W}_{2}(p){\rm d}^dp$ 
on the forward light cone ${\rm V}^+=\{ p\in \RR^d: 
p^{2}\geq 0, p_0\geq 0 \}$. We shall make use of the decomposition of 
${\cal H}^{(1)}$ as a 
direct integral
\beq
{\cal H}^{(1)} = \int^{\oplus} {\cal H}_{m}^{(1)}\,{\rm d}\rho(m^{2}),
\eeq
where ${\cal H}_{m}^{(1)} = L^{2}(\RR^{d}, M(p, 
m^{2})\Theta(p_{0})
\delta(p^{2}-m^2)d^{d}p)$. 

The smeared field operators $\Phi(f)=\int\Phi(x)f(x)dx$ are defined not only 
for 
test functions 
$f\in{\cal S}(\RR^d)$, but also for distributions
$f\in{\cal S}(\RR^d)'$ such that the Fourier transform $\tilde f$ belongs to 
the $L^2$ space with respect to the measure
$({\cal W}_{2}(p)+{\cal W}_{2}(-p)){\rm d}p$
on ${\rm V}={\rm V}^+\cup -{\rm V}^+$. 
For real $f$, the field operator $\Phi(f)$ is uniquely determined by the 
restriction 
$\tilde f\big|_{{\rm V}^+}$. It is a self adjoint operator on a natural domain 
in the Fock space and 
we may consider the unitary Weyl operators $W(f)=\exp{\rm i}\Phi(f)$. 
They satisfy the relation
\beq W(f)W(g)=e^{-K(f,g)/2}W(f+g)\label{weyl}\eeq
with
\beq K(f,g)=\int \left(\widetilde {\cal W}_{2}(p)-\widetilde {\cal 
W}_{2}(-p)\right)\tilde f(-p)\tilde g(p){\rm d}^dp.
\eeq
Moreover, 
\beq
\langle\Omega, W(f)\Omega\rangle=\exp\left(-\hbox{$\int$} 
\widetilde {\cal W}_{2}(p)
\tilde f(-p)\tilde f(p)dp\right)\label{weylexp}.
\eeq

If ${\cal O}$ is a subset of Minkowski space $\RR^{d}$ we 
can define the following subspace of ${\cal H}^{(1)}_{m}$
\beq\label{cc}
{\cal H}_{m}^{(1)}({\cal O}) := {\rm closure\ of\ }
\left\{ \tilde g\big|_{H_{m}^+}:\ g\in{\cal S}({\RR^d}),\ {\rm supp}\,g \subset 
{\cal O} \right\}.
\eeq
We define the local
algebra ${\cal M}({\cal O})$ as the von Neumann algebra generated by the  Weyl 
operators $W(f)$ (with real $f\in {\cal S}^1(\RR^d)'$) such that
\beq
\tilde f\big|_{{\rm V}^+}\in \int^{\oplus} {\cal H}_{m}^{(1)}({\cal O})\,{\rm 
d}\rho(m^{2}).
\eeq
We remark that if 
the Lehmann weight does not decrease rapidly at infinity then ${\cal M}({\cal 
O})$ 
can be larger than the algebra 
generated by the Weyl operators $W(f)$ with ${\rm supp\ } f\subset {\cal O}$, 
cf.\ \cite{L}. This possibility, however, is independent of the issues of 
interest here. Our definition of ${\cal 
M}({\cal O})$ simplifies things because it allows a complete reduction to the 
case of fixed mass.

If $\OO$ is a fixed open subset of $\RR^d$ such that its causal 
complement $\OO'$ has a nonempty interior, then $\Omega$ is cyclic 
and separating for ${\cal M}({\cal O})$ and we may consider the 
corresponding modular group $\Delta^{{\rm i}t}$ and modular conjugation 
$J$.  Both are the second quantization of their restrictions to the 
one-field space ${\cal H}^{(1)}$ and we denote these restrictions by 
by $\delta^{{\rm i}t}$ and $j$ respectively. Moreover, by our definition 
of ${\cal M}({\cal O})$, we have a direct integral decomposition of 
these objects:
\beq
\delta^{{\rm i}t} = \int^{\oplus} \delta_{m}^{{\rm i}t}\,{\rm 
d}\rho(m^{2}),\qquad j=\int^{\oplus} j_{m}\,{\rm 
d}\rho(m^{2}).
\eeq
Here $\delta_{m}^{{\rm i}t}$ and $j_{m}$ the restrictions
to the one-field space $\HH_{m}^{(1)}$ of the modular objects 
for
the field with
2-point-function  %
\beq\label{d}
\widetilde{\cal W}_{2,m}(p) = M(p, m^{2})\,\Theta(p_{0})\,\delta(p^{2} - 
m^{2}).
\eeq
It is therefore sufficient to compute the modular objects for a fixed 
mass and we shall in the sequel drop the index $m$. We shall also 
write $M(p, m^{2})$ simply as $M(p)$.

\section{Computation of the modular objects for wedge algebras}

We shall now compute $\delta^{{\rm i}t}$ and $j$ for the field with 
two point function (\ref{d}) and $\OO$ a space like wedge $W$. Since the 
field is translationally covariant and general polynomials $M$ are 
allowed in (\ref{d}) it is sufficient 
to do this for some standard wedge. We choose for this purpose 
the ``right wedge''
\beq 
W_{\rm R}=\{x=(x_{0},\dots,x_{d-1})\in\RR^d : |x_{0}|<x_{1}\}
\eeq
The modular objects to this wedge will be denoted 
$\delta_{{\rm R}}^{{\rm i}t}$ and $j_{\rm R}$.
If $\Lambda$ is a Lorentz transformation, then the modular objects 
for the wedge $W=\Lambda W_{\rm R}$ are the same as for $W_{\rm R}$, but  with 
the 
polynomial $M_{\Lambda}(p):=M(\Lambda^{-1}p)$ instead of $M$. 
 
We introduce the light cone coordinates $p_{\pm} := 
p_{0}\pm p_{1}$, and write the remaining components of $p$ as  
$\hat p := (p_{2}, \ldots ,p_{d-1})$. The two point function
(\ref{d}) can then be written as
\beqa
\widetilde{\cal W}_{2}(p)& = &
 M(p_{+}, p_{-}, \hat p)\,\Theta(p_{+})\,\delta(p_{+}\cdot p_{-} - 
\hat p^{2} - m^{2})\\ &=&
{p_{+}^{-1}} M(p_{+},p_{+}^{-1}({\hat p^{2} + m^{2}}) , \hat 
p)\,\Theta(p_{+})\,
 \delta\left(p_{-} - 
p_{+}^{-1}({\hat p^{2} + m^{2}})\right).
\eeqa
Moreover, since $M$ is a polynomial we can write
\beq
M(p_{+},p_{+}^{-1}({\hat p^{2} + m^{2}}),\hat p) =p_{+}^{-2n}Q(p_{+}, \hat 
p)
\eeq
with some $n\in\NN\cup\{0\}$ and a polynomial $Q(p_{+}, \hat p)$.  The 
properties of $M$ imply that $Q$ satisfies \beq Q(p_{+}, \hat p)= 
Q(-p_{+}, -\hat p)\quad{\rm and}\quad Q(p_{+}, \hat p)\geq 0.\label{prop} \eeq

We now consider $Q$ as a polynomial in $p_{+}$, with coefficients that 
are polynomials in $\hat p$.  Its zeros are algebraic functions of 
$\hat p$, and the properties (\ref{prop}) entail that every real zero 
$r_{j}(\hat p)$ of $Q(\cdot, \hat p)$ must be a double zero and every 
complex zero $z_{k}(\hat p)$ comes together with its complex 
conjugate $z_{k}(\hat p)^*$.  Moreover, each real zero 
$r_{j}(\hat p)$ is accompanied by a zero $-r_{j}(-\hat p)$ and every 
complex zero $z_{k}(\hat p)$ by $-z_{k}(-\hat p)$.

All in all we can write
\beq\label{e}
\widetilde{\cal W}_{2}(p) = \frac{1}{p_{+}}\,F(p_{+}, \hat 
p)\,F(-p_{+}, -\hat p)\,
\Theta(p_{+})\,
\delta\left(p_{-} - \frac{\hat p^{2}+m^2}{p_{+}}\right),
\eeq
with
\beq\label{f}
F(p_{+}, \hat p) = \frac{1}{(ip_{+})^{n}}\cdot\prod_{j=1}^{J}(p_{+} - 
r_{j}(\hat p))
(p_{+} + r_{j}(-\hat p))\prod_{k=1}^{K}(p_{+} + z_{k}(\hat p))(p_{+} - 
z_{k}(-\hat p)^{*}),
\eeq
where $r_{j}(\hat p) \in \RR$ and $z_{k}(\hat p) \in 
\CC $, ${\rm 
Im}\,z_{k}(\hat p) > 0$. 
Thus $F$ has all the complex zeros of $Q$ in the lower half plane and 
no zeros in the upper half plane, while 
\beq
F(-p_{+}, -\hat p) = F(p_{+}, \hat p)^{*}\label{real}
\eeq
has no zeros in the lower half plane. The real zeros of $Q$ are evenly 
divided between $F(p_{+}, \hat p)$ and $F(-p_{+}, -\hat p)$.

We shall now give  explicit formulas for $\delta^{\rm it}_{\rm R}$ and $j_{\rm 
R}$. 
Note that every $\varphi\in\HH^{(1)}_{m}$ can be regarded as a function 
of $p_{+}>0$ and $\hat p\in\RR^{d-2}$, since on the mass shell 
$p_{-}=p_{+}^{-1}(\hat p^2+m^2)$.
\begin{theorem}
On the one particle space $\HH^{(1)}$ the modular group associated with ${\cal
M}(W_{\rm R})$ and ${\Omega}$ has the form
\beq\label{g}
(\delta^{\rm it}_{\rm R}\varphi) (p_{+},\hat p) = \frac{F(e^{2\pi t}p_{+}, \hat
p)}{F(p_{+}, \hat p)} \varphi(e^{2\pi t}p_{+}, \hat
p)
\eeq
where $F$ is given by (\ref{f}). The corresponding modular conjugation 
is
\beq\label{h}
(j_{\rm R}\varphi)(p_{+},\hat p) = \frac{F(-p_{+}, \hat
p)}{F(p_{+}, \hat p)} \varphi(p_{+}, -\hat p)^{*}
\eeq
\end{theorem}
\begin{proof} One can easily check that $\delta^{\rm it}_{\rm R}$ is unitary 
for 
all
$t$ and $j_{\rm R}$ is anti-unitary. The same holds then for the second 
quantized operators $\Delta^{\rm it}_{\rm R}$ and $J_{\rm R}$. To show 
that $\Delta^{\rm it}_{\rm R}$ is indeed the 
modular group 
associated with the vacuum state on ${\cal M}(W_{\rm R})$ it is necessary to 
verify that $\sigma_{t}:={\rm ad}\Delta^{\rm it}_{\rm R}$ defines an 
automorphism group of ${\cal M}(W_{\rm R})$  and that the 
KMS  condition
\beq \langle\Omega, \sigma_{t}W(f)W(g)\Omega\rangle=
\langle\Omega, W(f)\sigma_{t-{\rm i}}W(g)\Omega\rangle\eeq
holds for Weyl operators localized in $W_{\rm R}$.

By Eq.\ (\ref{g}), (\ref{weyl}) and (\ref{weylexp}) the action of $\Delta^{\rm 
it}_{\rm R}$ on the Weyl 
operators is
\beq 
\Delta^{\rm it}_{\rm R}W(f)\Delta^{-it}_{\rm R}
=W(f_{t})
\eeq
with
\beq
\tilde f_{t}(p_{+},p_{-},\hat p)=\frac{F(e^{2\pi t}p_{+}, \hat
p)}{F(p_{+}, \hat p)}\tilde f(e^{2\pi t}p_{+}, e^{-2\pi t}p_{-},\hat
p).
\eeq
(Note that on the positive and negative mass shells $p_{+}p_{-}=\hat 
p^2+m^2$.) Test functions $f$ with support in $W_{{\rm R}}$ are 
characterized by analyticity and decay properties of the Fourier 
transform $\tilde f$: For fixed $\hat p$, $\tilde f$ is analytic in 
\beq
{\cal T}_{\rm R}=\{(p_{+},p_{-})\in \CC^2: {\rm Im\ } p_{+}>0,\
{\rm Im\ } p_{-}<0\}
\eeq
and decays rapidly at infinity in this domain. The same conditions
apply if $f$ 
is a distribution w.r.t.\ the light cone variables $x_{\pm}$, but 
$\tilde f$ may increase like an inverse polynomial as $p_{+}$ or 
$p_{-}$ approach the real axis. Since $F$ has no zeros in $p_{+}$ in 
the open upper half plane, it is evident that $f_{t}$ satisfies these 
conditions if $f$ does. Hence the group ${\rm ad}\Delta^{\rm it}_{\rm R}$ 
leaves 
$\MM(W_{\rm R})$ invariant.

The KMS condition can be verified by essentially the same 
computation as the corresponding statement for fields on a light ray 
in \cite{Y}.

To show that (\ref{h}) is the modular conjugation we note first that 
the set of state vectors
$\varphi\in\HH^{(1)}$, such that $\varphi=\tilde f\big|_{H^{+}_{m}}$ 
with $f\in{\cal S}(\RR^{d})$ and supp $f\in W_{\rm R}$, is a core for 
the restriction $s$ to $\varphi\in\HH^{(1)}$ of the $S$ operator corresponding  
to $\MM(W_{\rm R})$ and $\Omega$. The latter is defined by
$SW(f)\Omega=W(f)^{*}\Omega$ for supp $f\in W_{\rm R}$. Such 
$\varphi$ are  analytic in 
$p_{+}$ in the upper half plane, and
\beq
\delta^{1/2}_{\rm R}\varphi (p_{+},\hat p)=\frac{F(-p_{+}, \hat
p)}{F(p_{+}, \hat p)}\varphi(-p_{+},\hat p)
\eeq
by analytic continuation of (\ref{g}) to $t={-\rm i}\pi/2$. On the 
other hand, 
\beq
s\varphi(p_{+},\hat p)=\varphi(-p_{+},-\hat p)^{*}
\eeq
Using (\ref{real}) we see that $j_{\rm R}$ satisfies
\beq
s=j_{\rm R}\delta^{1/2}_{\rm R}
\eeq
as required for the modular conjugation.
\end{proof}

\section{Duality and modular action for a fixed wedge}

As next topic we discuss duality and the geometrical significance of the modular 
objects for the right wedge. In particular we compare them with the 
corresponding objects for the left wedge
\beq 
W_{\rm L}=\{x=(x_{0},\dots,x_{d-1})\in\RR^d : |x_{0}|<-x_{1}\}.
\eeq
By an analogous computation as for the right wedge these are given by
\beq\label{gl}
(\delta^{\rm it}_{\rm L}\varphi) (p_{+},\hat p) = \frac{F(-e^{-2\pi t}p_{+}, 
-\hat
p)}{F(-p_{+}, -\hat p)} \varphi(e^{-2\pi t}p_{+}, \hat p)
\eeq
and
\beq\label{hl}
(j_{\rm L}\varphi)(p_{+},\hat p) = \frac{F(p_{+}, -\hat
p)}{F(-p_{+}, -\hat p)}\varphi(p_{+}, -\hat p)^{*}.
\eeq
Wedge duality for the left and right wedge, i.e., $\MM(W_{\rm 
R})^\prime=\MM(W_{\rm L})$,
holds if and only if the modular 
conjugations coincide, i.e., $j_{\rm R}=j_{\rm L}$. By (\ref{h}) and (\ref{hl}) 
the condition for this is
\beq
\frac{F(-p_{+}, \hat
p)}{F(p_{+}, \hat p)} = \frac{F(p_{+}, -\hat
p)}{F(-p_{+}, -\hat p)},
\eeq
which by (\ref{real}) can be written
\beq
F(p_{+}, -\hat p) F(p_{+}, \hat p) = F(p_{+}, -\hat p)^* 
F(p_{+}, \hat p)^*.
\eeq
Since $F(p_+,\pm\hat p)$, regarded as a function of $p_{+}$, has 
all its complex zeros in the lower half plane, we see that this holds if and 
only if $F$ has no complex zeros at all.

Let us now consider the geometric action of the modular conjugation. 
If $M$ has only real zeros in $p_{+}$, then duality holds and hence $J_{\rm R}
\MM(W_{\rm R})J_{\rm R}= \MM(W_{\rm L})$.  A complex zero, 
on the other hand,  implies that the pre factor $F(- p_{+}, \hat
p)/F(p_{+}, \hat p)$ in the definition of $j_{\rm R}$ is not analytic in the 
lower half plane. Hence in general $j_{\rm R}\varphi$ is not analytic 
in $p_{+}$ in the lower 
half plane for $\varphi\in\HH^{(1)}(W_{{\rm R}})$.  This implies that 
$j_{\rm R}\varphi$
is in general not contained in $\HH^{(1)}(W)$ for any wedge of the form of 
$W_{\rm L}+a$, $a\in\RR^d$, and hence 
$J_{\rm R}
\MM(W_{\rm R})J_{\rm R}$ is not contained in any translate of $\MM(W_{\rm 
L})$. A localization of $J_{\rm R}
\MM(W_{\rm R})J_{\rm R}$ in any other wedge algebra is excluded since 
for general $\varphi\in\HH^{(1)}(W_{{\rm R})}$, $\varphi(p_{+},-\hat p)^{*}$ 
has 
no further analyticity properties beyond the analytic continuation in 
$p_{+}$ to the lower half plane which follows from the localization 
of $\varphi$ in $W_{{\rm R}}$. 

We summarize these findings as follows.
\begin{proposition}
The following are equivalent
\begin{itemize}
\item[(i)] $\MM(W_{\rm 
R})^\prime=\MM(W_{\rm L})$
\item[(ii)] $J_{\rm R}\MM(W_{\rm 
R})J_{\rm R}$ is contained in $\MM(W)$ for some wedge $W$.
\item[(iii)] The rational function 
\beq p_{+}\mapsto M(p_+,p_+^{-1}(\hat p^2+m^2), \hat p)\eeq 
has only real zeros, for all $\hat p\in\RR^{d-2}$.
\end{itemize}
\end{proposition}

Our last concern in this section is the local action of the modular group. 
The general theorem of Borchers \cite{B2} implies that translates of 
$\MM(W_{\rm R})$  are mapped onto algebras of the same type:
\beq \Delta_{\rm R}^{\rm it}\MM(W_{\rm R}+a)\Delta_{\rm R}^{-\rm it}
=\MM(W_{\rm R}+\Lambda(t)a)\eeq
for all $a\in\RR^d$, with $\Lambda(t)$ a Lorentz boost. Observables 
localized in bounded domains, however, are in general not localized in a 
bounded domain after transformation by
${\rm ad}\Delta_{\rm R}^{\rm it}$. In fact, if $\OO$ is bounded, then
$\varphi\in\HH^{(1)}(\OO)$ is the restriction to the mass shell of an 
entire analytic function. This analyticity is in general destroyed by the pre 
factor $F(e^{2\pi t} p_{+}, \hat p)/F(p_{+}, \hat p)$, unless  
$F(p_{+},\hat p)$ and $F(e^{2\pi t} p_{+},\hat p)$ have the same 
set of zeros. This holds only if $M(p_+,p_+^{-1}(\hat p^2+m^2),\hat p)$ has the 
form $p_{+}^{2n}C(\hat p)$ for some $n\in{\mathbb Z}$. If 
$M$ is independent of $\hat p$, then there is at least no 
dislocalization in the directions along the edge of the wedge, but 
the example $M(p)=p_{0}^2$ mentioned in {\cite{Y}} (this corresponds 
to the time derivative of the free field) has 
$F(p_{+},\hat p)=(2\rm ip_+)^{-2}(p_{+}+\rm i(\hat p^2+m^2)^{1/2})^2$ 
and here $F(e^{2\pi t} p_{+}, \hat p)/F(p_{+}, \hat p)$ also 
dislocalizes in the $\hat x$ variables if there are such variables at 
all, i.e., if $d\geq 3$.

\section{Duality and modular action for all wedges}

In the last section we dealt with a fixed wedge and 
saw in particular that duality for $W_{\rm R}$ and $W_{\rm L}$ holds if 
and only if $M$ has only real zeros in $p_{+}$ on the mass shell. For 
$d=2$ this is the complete answer to the question when wedge duality holds
and this does not necessarily imply Lorentz covariance of the 
field. 

We shall now see how the picture changes in dimensions $d\geq 3$. We start with 
the local action of the modular groups.
\begin{proposition}\label{local}
Suppose $d\geq 3$ and the modular group for every wedge acts locally on the net 
generated by the field. Then $M$ is constant on the mass shell.
\end{proposition}
\begin{proof}
By the discussion in the last section local
action of the modular group $\Delta_{\rm R}^{\rm it}$ requires that
$M(p_+,p_+^{-1}(\hat p^2+m^2),\hat p)$ has the form $p_{+}^{2n}C(\hat p)$ for
some $n\in{\mathbb Z}$ and a function (polynomial) $C$ depending only on 
$\hat p$.  If  $M_\Lambda$  has the same form for all $\Lambda$ then in
particular we have for the Lorentz boosts $\Lambda_W(t)$ corresponding to an 
arbitrary wedge $W$ and boost parameter $t$ 
\begin{equation} M(\Lambda_W(t)^{-1}p)=D(\Lambda_W(t))M(p) \end{equation} with
$D(\Lambda_W(t))=(\exp(2\pi t))^{2n_W}$ for some $n_W\in{\mathbb Z}$.  Moreover, 
since this holds for all $p$ on the mass shell,
we conclude that 
$D(\Lambda_{W_1}(t)\Lambda_{W_2}(s))=D(\Lambda_{W_1}(t))D(\Lambda_{W_2}(s))$ for 
any two boosts in arbitary directions.  Since any Lorentz 
transformation can be written as
a product of boosts, we obtain in this way a one dimensional representation of
the Lorentz group.  If $d\geq 3$ this implies that $D$ is constant, and hence,
since the Lorentz group acts transitively on the mass shell, that $M$ is
constant on the mass shell.
\end{proof}

The requirement that wedge duality 
holds for {\it all} wedges also restricts the possible structure of $M$ 
drastically in higher dimensions than 2. This is due to the 
following
\begin{lm}\label{basic} Let $M(p_+,p_-,\hat p)$ be an even polynomial on 
$\RR^d$ with $d\geq 4$.
If the rational function \beq p_+\mapsto M_\Lambda(p_+,p_+^{-1}(\hat
p^2+m^2),\hat p)\eeq has only real zeros for every Lorentz
transformation $\Lambda$ and every $\hat p\in\RR^{d-2}$, then $M$ is 
constant on
the mass shell $H^+_m$.  The same holds for $d=3$ if $m>0$.  \end{lm}
\begin{proof} We denote the rational function 
$M(p_+,p_+^{-1}(\hat
p^2+m^2),\hat p)$ by $R(p_+,\hat p)$ for short. If $\Lambda$ is a Lorentz 
transformation, then the passage from $M$ to $M_\Lambda$ replaces $R(p_+,\hat 
p)$ by $R_\Lambda(p_+,\hat p)=R((\Lambda^{-1} p)_+, {(\Lambda^{-1} 
p)}{\hat{}}\,)$. Suppose now that $R$ is 
{\it not} constant. Since $\Lambda$ is invertible it is clear that $R_\Lambda$ 
is not constant either for any $\Lambda$. We shall show that there exists a 
Lorentz transformation $\Lambda$ and a $\hat p\in\RR^{d-1}$ such that 
$p_+\mapsto R_\Lambda (p_+,\hat p)$ has a complex (i.e. not real) zero. 

The function $R$ has the form
\begin{equation}
R(p_+,\hat p)=\sum_{n\in{\mathbb Z}}p_+^na_n(\hat p)
\end{equation} 
where the $a_n$ are polynomials in $\hat p$, and $a_n\equiv 0$ except for 
finitely 
many $n$. 
Likewise,
\begin{equation}
R_\Lambda(p_+,\hat p)=\sum_{n\in{\mathbb Z}}(\Lambda^{-1}p)_+^na_n(\Lambda^{-1} 
p{\hat{}}\,)=\sum_{n\in{\mathbb Z}}p_+^na_n^\Lambda (\hat p)
\end{equation} 
with different coefficients $a_n^\Lambda (\hat p)$. The first remark is that 
there is at least one $n\neq 0$  such that $a_n^\Lambda$ is not 
identically zero for some $\Lambda$. In fact, suppose $R$ is independent of 
$p_+$, i.e.,
$R(p_+,\hat p)=a_0(\hat p)$. Since the polynomial $a_0$ is not constant by 
assumption, it 
depends nontrivially on $p_i$ for at least one $i$, $2\leq i\leq d-1$, i.e., it 
contains a term $p_i^\nu b_\nu(p_2,\dots,p_{i-1},p_{i+1},\dots p_{d-1})$ with 
$\nu\neq 0$. If $\Lambda$ is a rotation by $\pi/2$ in the $1i$ plane, 
then
$(\Lambda^{-1} p)_i=p_1$. For $p$ on the mass shell
\beq
p_1=\mfr1/2(p_+-p_-)=\mfr1/2(p_+-(\hat p^2+m^2)p_+^{-1})
\eeq 
and inserting this for $(\Lambda^{-1} p)_i$ we see that $R_\Lambda$ is not 
independent of $p_+$. To simplify 
notation we denote this $R_\Lambda$ again by $R$.

Since we may now assume that $R$ depends nontrivially on $p_+$, we can write
\beq
R(p_+,\hat p)=p_+^{-2n}A(\hat p)B(p_+,\hat p)
\eeq
where $A(\hat p)$ is a polynomial and
\beq 
B(p_+,\hat p)=p_+^{2\ell}+\hat B(p_+,\hat p)
\eeq
with $\ell\geq 1$ and $\hat B(p_+,\hat p)$ a polynomial in $p_+$ of degree lower 
than 
${2\ell}$.
The coefficients of this polynomial are  
real analytic functions of $\hat p$ on some open set in $\RR^{d-1}$. We write 
$\hat p=(p_2,\tilde 
p)$ with $\tilde p\in\RR^{d-3}$ (if $d=3$ there is no $\tilde p$) and fix 
$\tilde p$. Then $B$ can be regarded as a polynomial in $p_+$ with coefficients 
that are real analytic in $p_2$ on some open interval. The coefficient to the 
highest power of $p_+$ is independent of $p_2$.

If $B$ has a complex zero in $p_+$ for some $p_2$ there is nothing more to be 
proved.  On the other hand, if all zeros of
$B$ are real we may apply a theorem of Rellich
\cite{R} (see also \cite{AKLM}), from which it follows that there is a
real analytic function $r(\cdot)$, so that $p_+=r(p_2)$ is a zero of
$B$, and hence of $R$, for all $p_2$ in some open interval. Since $M$ and hence 
$R$ is even, we may assume that $r(p_2)> 0$. 
(The case $r(p_2)\equiv 0$ would mean that $M$ on the mass shell has the form 
$p_+^{2n}C(\hat p)$. As shown in Proposition \ref{local} this can not hold in 
all Lorentz systems unless $M$ is constant on the mass shell.)

It is convenient to replace the
variables $(p_+,p_2,\tilde p)$ on the mass shell by the variables
$(p_1,p_2,\tilde p)$: \beq
p_1=\mfr1/2(p_+-p_-)=\mfr1/2(p_+-(p_2^2+\tilde
p^2+m^2)p_+^{-1}).\label{p1}\eeq The inverse transformation is \beq
p_+=p_0+p_1=(p_1^2+p_2^2+\tilde p^2+m^2)^{1/2}+p_1.\label{p+}\eeq
Inserting $p_+=r(p_2)$ in (\ref{p1}) we obtain a real analytic curve
\beq p_1=s(p_2)\label{curve}\eeq of zeros of $B$, and hence of $R$, in
the $12$-plane. 

The function $R_\Lambda$ has a corresponding curve of zeros at fixed $\tilde p$ 
for any Lorentz transformation $\Lambda$ that affects only the variables 
$p_0,p_1$ and $p_2$. This curve is given by $(\Lambda p)_+=r((\Lambda p)_2)$, or 
equivalently in the variables $p_1,p_2$, by $(\Lambda p)_1=s((\Lambda p)_2)$. 
The point $p\in\RR^d$ is here always on the mass shell.

Returning to the original curve  $p_1=s(p_2)$ there are two possibilities:
\begin{itemize}
\item The curve is a straight line segment.
\item There is a point $\bar p_2$, such that the second derivative 
$s^{\prime\prime}(\bar p_2)\neq 
0$.
\end{itemize}
We deal with the second case first. 

By Taylor expansion we have 
\beq 
s(p_2)=s(\bar p_2)+s'(\bar p_2)(p_2-\bar p_2)+\mfr1/2s^{\prime\prime}(\bar 
p_2)(p_2-\bar p_2)^2(1+g(p_2-\bar p_2))
\eeq
whith some real analytic function $g$ satisfying $g(t)\to 0$ for $|t|\to 0$. Let 
$\Lambda$ be a rotation in the 12 plane by 
an angle $\varphi$, determined by $\cot\varphi= s'(\bar p_2)$. This  
transformation rotates the curve so that the tangent which previously had the 
slope $s'(\bar p_2)$ becomes parallel to the 1-axis. Moreover, the point 
$(s(\bar p_2),\bar p_2)$ is rotated into another point, $(a,b)=(\cos\varphi\, 
\bar p_2+\sin\varphi\, s(\bar p_2), -\sin\varphi\, \bar p_2+\cos\varphi\, s(\bar 
p_2))$, while 
the curvature, $\mfr1/2s^{\prime\prime}(\bar p_2)=:c\neq 0$ remains unchanged. 
Hence 
the equation of the the rotated curve, i.e.,
$(\Lambda p)_1=s((\Lambda p)_2)$, has the form
\beq p_2=b+c(p_1-a)^2(1+h(p_1-a)),\label{neweq}\eeq
with $a,b,c\in\RR$, $c\neq 0$ and where $h$ is real analytic with $h(t)\to 0$ 
for $t\to 
0$. 

By analytic continuation, $R_\Lambda$ vanishes also for complex points $p_1$
satisfying this equation.  It is clear that if $(p_2-b)/c$ is negative and 
sufficiently 
small,
then there is a solution for $p_1$ with a nonvanishing imaginary part.  By 
Eq.\ 
(\ref{p+}) this corresponds to a $p_+$ with nonvanishing imaginary part. (Note 
that $p_2$ and $\tilde p$ are still real.)
Hence $R_\Lambda$ has a complex zero in $p_+$ for some 
$(p_2, \tilde p)\in\RR^{d-1}$.

If the curve (\ref{curve}) is a straight line, we can by a rotation transform it
to a line parallel to the $p_1$ axis, \beq p_2=k\label{line}\eeq with a constant
$k$.  A Lorentz boost in the $2$-direction with parameter $\alpha$ transforms
(\ref{line}) into \beq p_2=(\cosh \alpha)k+(\sinh\alpha)(k^2+p_1^2+\tilde
p^2+m^2)^{1/2}.\label{lotr}\eeq If $k^2+\tilde p^1+m^2>0$ we are back to the
case considered before.  This can always be achieved by choosing $\tilde p\neq
0$ if $d\geq 4$, and it holds also for $d=3$ if $m>0$.  Thus we have again found
a $\Lambda$, this time a composition of a rotation and a Lorentz boost, such
that $R_\Lambda$ has a complex zero in $p_+$.  
\end{proof}

The following examples show that wedge duality and Lorentz invariance are not
necessarily equivalent in lower dimensions than 4.  
\medskip

\noindent{\large{\bf Examples}}
\medskip

1. Consider a massless field in $d=3$ with $M(p)=(a\cdot p)^{2n}$ where 
$a=(a_0,a_1,a_2)$ is a 
space-like, or light like vector in $\RR^3$. (The exponent $2n$ guarantees 
the 
required positivity and symmetry.) It is clear that $M_\Lambda$ has the 
same 
form for all Lorentz transformations $\Lambda$.
Vanishing of $M$ is the same as vanishing of $a\cdot p$, and on the mass 
shell 
\begin{eqnarray} a\cdot 
p&=&\mfr1/2a_0(p_++p_2^2p_+^{-1})-\mfr1/2a_1((p_+-p_2^2p_+^{-1})-a_2p_2
\nonumber
\\ &=&p_+^{-1}
\left[\mfr1/2(a_0-a_1)p_+^2-(a_2p_2)p_++\mfr1/2(a_0+a_1)p_2^2\right].
\end{eqnarray}
The discriminant of the quadratic equation for $p_+$ is
\beq (a_2p_2)^2-(a_0-a_1)(a_0+a_1)p_2^2=-(a\cdot a)p_2^2\geq 0\eeq
for all real $p_2$, if $a$ is space-like or light-like. Thus there are only 
real 
zeros.
But $M$ is not constant on the mass shell, unless $a=0$.

\medskip

2. In $d=2$ we may also consider mass $m>0$ (fields without a mass gap, 
depending only on one light cone coordinate, are discussed in \cite{Y}): 
With 
$M$ as above we have on the mass shell
\begin{eqnarray} a\cdot 
p&=&\mfr1/2a_0(p_++m^2p_+^{-1})-\mfr1/2a_1((p_+-m^2p_+^{-1})\\ 
&=&\mfr1/2p_+^{-1}
\left[(a_0-a_1)p_+^2+(a_0+a_1)m^2\right].
\end{eqnarray}
Again, if $a$ is space-like or light like there are only real zeros. 
\medskip

\noindent{\it Remark.} In both these examples the minimal net of von
Neumann algebras, $\MM_{\rm min}(\OO)=\{W(f):{\rm supp\
}f\subset\OO\}^{''}$ generated by the field is different from the
maximal net $\MM_{\rm max}(\OO)=\MM_{\rm min}(\OO')'$, if $\OO$ is
bounded. However, for every wedge $W$ we have $\MM_{\rm
min}(W)=\MM_{\rm max}(W)$, and $\MM_{\rm max}(\OO)=\cap_{W\supset
\OO}\MM_{\rm min}(W)$. (This is a well known consequence of wedge
duality, cf.\, e.g., Lemma 4.1 in \cite{BY}). Moreover, $\MM_{\rm
max}(\cdot)$ is Lorentz covariant in both examples. In fact, it is
straightforward to verify (cf.\ Section 3 in \cite{Y}) that for $a$
space-like or light like, the Lorentz covariant field $\Phi_0$ and the
non-Lorentz covariant derivatives $a\cdot\partial \Phi_0$ generate the
same wedge algebras. In particular, CGMA also holds in these examples,
because the wedge algebras are generated by a Lorentz covariant
field. 
\medskip

Putting everything together we finally obtain the main conclusion of this note:
\begin{thm}
If $d\geq 4$ the following are equivalent for the generalized free
field models considered
\begin{itemize}
\item[(i)] CGMA
\item[(ii)] Wedge duality for all wedges
\item[(iii)] Local action of the modular groups of all wedges
\item[(iv)] Lorentz covariance of the field
\end{itemize}
For models with a mass gap this holds also for $d=3$. 
\end{thm}

\bigskip
\noindent{\large{\bf Acknowledgements}} 
\smallskip

We are grateful to H.J. Borchers and to P.W.\ Michor for very helpful 
discussions concerning Lemma \ref{basic}.

\end{document}